\documentclass[structabstract]{aa}
\usepackage{graphicx}
\usepackage{txfonts}
\usepackage{float}
\usepackage{perpage}
\MakeSorted{figure}
 \MakeSorted{table}

\begin{document}
   \title{Very Early Photometry of SN 1998S: Physical Parameters and Date of Explosion}


   \author{Poon, H.
          \inst{1}
          \and
          Pun, J. C. S.\inst{1}
           \and
          Lam, T. Y.\inst{2}
            \and
          Qiu, Y. L.\inst{3}
            \and
          Wei, J. Y.\inst{3}
          }

   \institute{Department of Physics, University of Hong Kong,Pokfulam Road, Hong Kong, PR China
              \\
              \email{china\_108@yahoo.com}
         \and
             Institute for the Physics and Mathematics of the Universe, University of Tokyo, Kashiwa, Chiba 277-8583, Japan\\
             \and
             National Astronomical Observatories, Chinese Academy of Science, 20a Datun road, Chaoyang District, Beijing, PR China\\
             }



  \abstract
   {We present very early optical lightcurves beginning 10 days before maximum of the Type IIn supernova
   1998S, covering the first four months after discovery. }
   {We examine the light evolution and try to compare the
   lightcurves to two analytical models(Nakar \& Sari(2010) and Rabinak
  \& Waxman(2011)) for a red supergiant star.}
   {The photometry was carried at the 60-cm telescope of the Xinglong Station of China. Broadband
filters Johnson$'$s \emph{B}, \emph{V} and Cousins$'$ \emph{R} were
used.}
   {The magnitude rose for the first few days and then dropped slowly afterwards. The two different
   models we use
   can fit the early lightcurves very well. The explosion date derived from the models is within the range
   of 1998 March 1.34 - 2.64(JD 2450873.84 - JD 2450875.14.) The radius of the progenitor is found
   to be $\sim$ 300 $R_{\bigodot}$ and $\sim$ 2000 $R_{\bigodot}$ for the model of Nakar \& Sari(2010) and Rabinak
  \& Waxman(2011) respectively. The constraint on mass and energy is not strong. The ranges of these two parameters are within that of a red supergiant.  }
   {}


   \maketitle
%

\section{Introduction}
SN 1998S was discovered by BAO on March 2.68, 1998 (Li \& Li 1998).
This supernova lies in NGC 3877. There was no sign of supernova on
the CCD image taken on Feb 23.7(Leonard et al. 2000). It is one of
the brightest Type Iln supernova observed so far.

Type II supernovae are characterized by the presence of hydrogen in
the spectrum. According to the shape of light curves, they can be
further classified as II-P and II-L (Barbon et al. 1979). The former
show a plateau while the latter show a linear decay. Schlegel(1990)
defined a new subclass of Type IIn. Their spectra show narrow
emission lines with weak or no P - Cygni absorption component.

Early-time lightcurves are especially important in which they can be
used to test different models to constrain the progenitor's
parameters. Modjaz et al. (2009) compared their early-time
bolometric lightcurves of 2008D to the model of Waxman et al. (2007)
and also that of Chevalier \& Fransson (2008). They estimated the
radius to be 1.2 $\pm$ 0.7 R$\odot$ for the former model and 12
$\pm$ 7 R$\odot$ for the later one. Gezari et al. (2010) applied the
model of Nakar and Sari(2010, hereafter  NS10) and RW11(2011,
hereafter RW11) to the early($<$ 3 d) UV/optical lightcurves of SN
IIP 2010aq. NS10 can fit the lightcurves well with an offset of -1.5
mag. Combining all the constraints, they found the radius of the
progenitor to be 700 $\pm$ 200 R$\odot$. For RW11, they fit the
lightcurves with
 parameters within the constraints. The fit agrees with the data
 well at the time of the first detection of the source, but
 the temperature evolution is slower.

Filippenko \& Moran (1998) first presented the spectra of SN 1998S
and found it to be a type II supernova. Due to the presence of
narraow H $\alpha$ emission lines(Filippenko \& Moran 1998), they
suggested that the supernova belongs to the type Iln. From the
\emph{BV} light curves and spectroscopic behavior, Liu et al. (2000)
indicated a similarity between the supernova and Type IIL SN 1979C.
Leonard et al. (2000) found a high degree of linear
 polarization in their spectropolarimetric observations which implied asphericity for the continuum-scattering
 medium. Chandra observations(Pooley et al. 2002) of the supernova at the age of three show overabundant
 heavy elements. Pozzo et al. (2004) presented late-time near-infrared
  and optical spectra of the supernova. From the shape and evolution
  of the H$\alpha$ and He I line profiles, they noticed a powerful
  interaction with a RSG wind. The variable component found in
  optical/UV spectra indicates slow-moving circumstellar outflows originating from
the red supergiant progenitor(Bowen et al. 2000).


 Fassia et al. (2000) presented contemporaneous photometry in the
optical and infrared bands between day 11 to 146 after discovery and
reported and a high IR excess at day 130. The result was interpreted
as due to thermal emission from dust grains in the vicinity of the
supernova. Liu et al. (2002) observed the supernova in the
\emph{BVR} bands and found the light curves typical of a SN II-L.

In this paper, we present optical photometry of SN 1998S spanning
the first four months. We reanalyze the data from Liu et al. (2000)
using the same data reduction method. We extend the early light
curves of SN 1998S by adding a new data point taken on the day of
discovery. We compare the new lightcurves to two models proposed by
 NS10 and RW11. These two models are for early
supernovae lightcurves. We have enough data to use model fits to
constrain the parameters and the date of explosion of the supernova.




\section{Observations and Results}
Our photometric observations were carried out at the 60 cm telescope
located at the Xinglong Station of the National Astronomical
Observatories of China (NAOC). The TI 215 CCD camera has 1024
$\times$ 1024 square pixels, a field of view of 16.$'$8 $\times$
16.$'$8, a gain of 11.6 e$^{-1}$/ADU and readout noise of 8
e$^{-1}$. The exposure time was about 120 s when the SN was around
maximum and 300 s when it was dim. For the first 10 days after
discovery, there are 6 data points taken, with the first data point
taken on the day of discovery(1998 March 2). Altogether, data of 32
nights were obtained, spanning four months. Standard \emph{BVR}
filters were used except for the discovery data point, which was
unfiltered.

After flat-fielding, bias, and dark corrections, aperture photometry
was performed using the apphot task of IRAF. The comparison star we
use is GSC 3452-1061. Its magnitudes are \emph{B} = 13.06 $\pm$
0.03, \emph{V} = 12.57 $\pm$ 0.007 and \emph{R} = 12.28 $\pm$
0.008(Fassia et al. 2000). For the unfiltered data point, we use the
following CCD transformation formula provided by BAO(private
communication):

\begin{equation}
i = I + i_{1} + i_{2}X_{i} + i_{3}(V - I) + i_{4}X_{i}(V - I)
\end{equation}

where \emph{i} is the instrumental unfiltered magnitude, \emph{I} is
the standard unfiltered magnitude and \emph{V} is the standard
\emph{V} band photometry. As differential photometry would be
applied, the values of the coefficients \emph{i$_{1}$} and
\emph{i$_{2}$} are neglected and \emph{i$_{3}$} = -0.039 $\pm$ 0.021
and \emph{i$_{4}$} = 0. Then equation (1) becomes
\begin{equation}
i_{SN} - i_{star} = I_{SN} - I_{star} + i_{3}(V_{SN} - I_{SN}) -
i_{3}(V_{Star} - I_{Star})
\end{equation}
We take the unfilterd magnitude as the standard \emph{R} magnitude
by approximation. Since the value of i$_{3}$ is small and the
comparison star has similar \emph{V} and \emph{R }magnitudes, the
last term \emph{i}$_{3}$(\emph{V$_{Star}$} - \emph{I$_{Star}$})
$\approx$ 0.01 and is neglected. To estimate the value of this term,
\emph{i}$_{3}$(\emph{V}$_{SN}$ - \emph{I}$_{SN}$), we first
calculate the temperatures of the supernova by converting our \emph{
BVR} data to fluxes and then perform a blackbody fit. The results
and the errors are listed in the last column of table 1. We then
extrapolate the temperature by performing an exponential fit and
find the temperature to be 27300 + 12200/-10500 K for the unfiltered
data point. The results are shown in fig. 1. By assuming a blackbody
spectrum, at T = 27300 + 12200/-10500 K, the magnitude difference
between \emph{V} and \emph{R} bands is $\sim$ 0.18 - 0.27. Thus, the
term \emph{i}$_{3}$(\emph{V}$_{SN}$ - \emph{I}$_{SN}$) is $\sim$
0.01 at maximum and can be neglected. We also try to determine this
temperature by performing a 3rd order polynomial. A result of 25000
$\pm$ 8900 K is found. Altogether, the error in neglecting the last
two terms of equation (2) should not be more than 0.02, plus the
error of the instrumental magnitude given by IRAF, which is 0.09,
the final error of the unfiltered data point is $\sim$ 0.11. The
photometry for all bands is listed in Table 1. The numbers in the
brackets represent the errors. Figure 2 shows the lightcurves in the
\emph{BVR} bands. Constants are added to individual light curve. The
photometry data from Fassia et al. (2000) are included for
comparison.

\begin{figure}[!ht]
   \centering
   \includegraphics[angle=90,width=7cm]{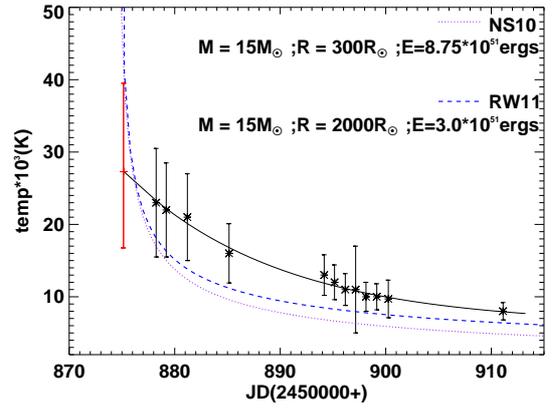}
      \caption{Exponential fit(black solid line) of the temperatures
of 1998S. The red dot represents the extrapolated temperature of the
discovery data point. The purple dotted line represents the
analytical model of NS10 and the blue dashed line represents that of
RW11. The best fitting parameters of these two models are used.
              }
         \label{FigVibStab}
   \end{figure}

\begin{figure}[!ht]
   \centering
   \includegraphics[angle=90,width=7cm]{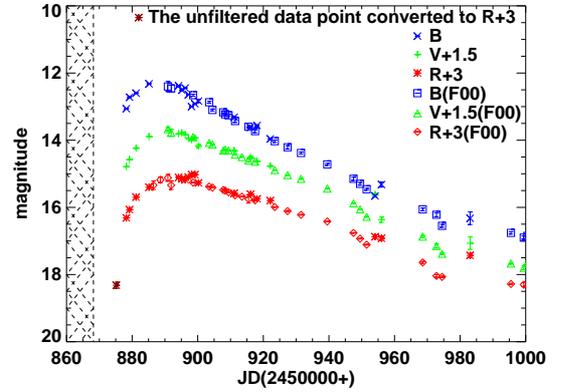}
      \caption{ \emph{BVR} light curves of SN 1998S. Data from Fassia
      et al.(2000)(F00) are included for comparison. The dashed
      straight line on JD = 2450868.2 indicates the earliest possible date
      of explosion.
                 }
         \label{FigVibStab}
   \end{figure}

%
\section{Comparison with Analytical Models}

There are different models which can explain the lightcurves of
supernovae. In the 80s, Arnett (1980, 1982) developed an analytical
model of radioactive decay diffusion. In this model, the radioactive
decay of nickel and cobalt is the main power source which diffuses
out from the expanding envelope. This model can successfully model
the late(t = 20 - 1500 days)bolometric lightcurves of 1987A(Arnett
\& Fu 1989). Later in the 90s, Chugai(1992) suggested a shock-wind
interactions model for type II supernovae at a late stage. The shock
interaction of the supernova shell with the wind of the presupernova
is the main source of energy for this model. The anomalous late-time
lightcurves of SN 1987F and SN 1988Z can be duplicated in the model.
More recently, Chevalier \& Fransson(2008) proposed a simple model
for the optical/ultraviolet emission from shock breakout. Since this
model requires a small radius progenitor, it is applicable to Type
Ib/c supernovae.

We compare our lightcurves to two analytical models for an RSG for
early supernova lightcurves and try to constrain the progenitor's
parameters. NS10 takes into account photon-gas coupling and
deviations from thermal
 equilibrium. The breakout is observed and then rises to a maximum luminosity in a very short time($<$ 1 day).
Immediately after the breakout, the luminosity keeps rising until a
diffusion time t$_{0}$(this is the luminosity of a radiation that
leaks from the center of a static slab). Then after t$_{0}$, light
travel time effects are included and the luminosity keeps on rising
and reaches a maximum value at R$_{star}$/c. At R$_{star}$/c,
  the lightcurve starts to decay in the planar stage(i.e., before the expanding gas doubles its radius) and then it
  rises again at the transition between the planar and spherical
  phase. The luminosity then drops until the frequency range reaches
  its maximum in the blackbody spectrum. The effects of recombination are not included, which is considered to be important only at a later
stage. RW11 focuses on the UV/O emission around $\sim$ 1 day
following the X-ray outburst. The breakout phase is not considered.
This model extends the analytic model of Waxman et al. (2007) to
include an approximate description of the time dependence of the
opacity(due mainly to recombination), and of the deviation of the
emitted spectrum from a black body spectrum. RW11 applied their
model to the early UV/O lightcurves of the type Ib SN2008D and of
the type IIP SNLS-04D2dc and found the results consistent with the
predictions of this model.

\subsection{Date of Explosion}
We first try to determine the date of explosion using both models.
Figure 3 shows the results of the best fits with different explosion
dates. We only consider fitting our lightcurves up to day 10 since
both models are for early lightcuvres and do not take into account
the production of nickel as a source of power supply at a later
stage. From the figure, the explosion date JD 2450874.34(1998 March
1.84) gives the best fit for NS10. The results of the fit with
t$_{exp}$ = JD 2450873.84 and JD 2450874.84 are still satisfactory,
only with considerable deviation from the first data point. For
RW11, t$_{exp}$ = JD 2450874.64(1998 March 2.14) gives the best fit.
t$_{exp}$ = JD 2450874.14 shows some deviation and t$_{exp}$ = JD
2450875.14 cannot fit the data at all. Hence, we conclude that the
explosion date is within the range of JD 2450873.84 - JD
2450875.14(1998 March 1.34 - 2.64). The first data point was taken
on day 0.69(taking the explosion date to be JD 2450874.49, the
average of the best results derived from both models) after
explosion. From the consistency between model predictions and the
evolution of photospheric radius, Chugai (2001) assumed the date of
explosion to be 1998 February 24.7 UT. He calculated the bolometric
luminosity, radius and velocity of the thin shell of the supernova
using a combined model which includes a model with ejecta-wind
interaction(Chugai 1992) and one
 without wind(Arnett 1980;1982). For the early data(before $\sim$ day 20) of the bolometric
luminosity(Fassia et al. 2000, the authors derived the bolometric
luminosity by performing a blackbody fit to their \emph{UBVRIJHK}
data), the analytical lightcurves even make a better fit under the
new date of explosion. Originally, the radius of the thin shell
derived from this model can fit the empirical radii well(Fassia et
al. 2000) but this result is a little worse if the date of explosion
is shifted backward by $\sim$ 5 days, . As for the velocity of the
thin shell, this is not affected at all since the model predicts a
steady value.

    \begin{figure}[!ht]
     \centering
     \includegraphics[angle=90,width=7.0cm]{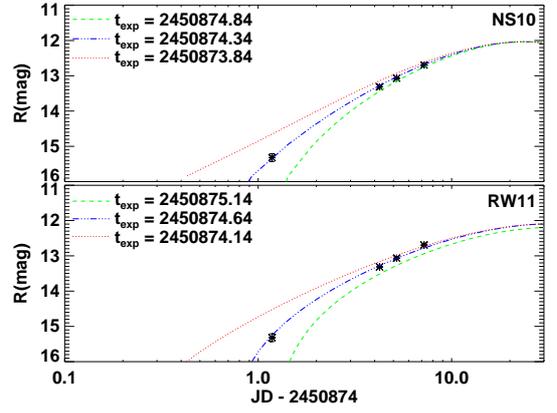}
        \caption{The best fit of NS10(top panel with E = 8.75$\times$10$^{51}$ ergs, M = 15$M_{\bigodot}$, R = 300$R_{\bigodot}$) and RW11(bottom panel with E = 3.0$\times$10$^{51}$ ergs, M = 15$M_{\bigodot}$,R = 2000$R_{\bigodot}$) for different explosion
        dates.
                        }
           \label{FigVibStab}
     \end{figure}
\subsection{Parameters of the Progenitor}

Figure 4 shows the results of the best fit for NS10 and RW11 for an
RSG with different radii and the best corresponding combinations of
mass and energy. The residuals are listed in Table 2. 
The first four data points were taken in the very early phase and
can very well constrain the parameters. All fits are satisfactory up
to day $\sim$ 6 for both models. For NS10, R = 300$R_{\bigodot}$, M
= 15$M_{\bigodot}$ and E = 8.75$\times$10$^{51}$ ergs can fit the
data up to day $\sim$ 25. For smaller radius, the luminosity first
drops to a lower one and then rises less sharply compared with a
larger radius. For RW11, R = 2000$R_{\bigodot}$, M =15$M_{\bigodot}$
and E = 3.0$\times$10$^{51}$ ergs fit all the data best. The
luminosity of the lightcuvres of this model all show similar
tendency in the very early phase but then it rises more sharply for
larger radii at a later stage.

\begin{figure}[!ht]
   \centering
   \includegraphics[angle=90,width=7cm]{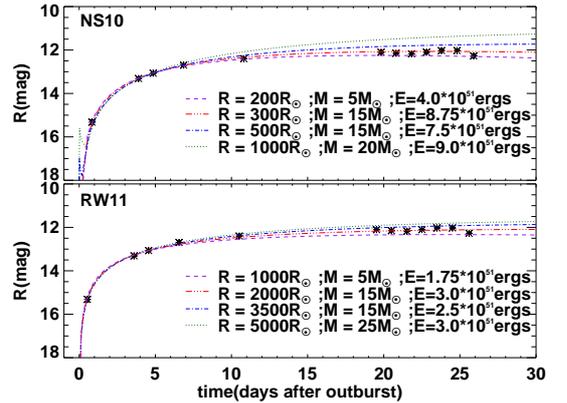}
      \caption{Zoom in of the optical lightcurves of SN 1998S in comparison to the analytical model
      of NS10(top panel) and RW11(bottom panel) with different radii and the best corresponding combinations of mass and energy. The dates of explosion for NS10
      and RW11
       in
      the figure are JD 2450874.34 and JD 2450874.64 respectively.
              }
         \label{FigVibStab}
   \end{figure}

   We try to investigate the importance of radius in both models by
   using different radii and fixed mass and energy to fit the data. In figure 5, we show the lightcurves for NS10 and
   RW11
with fixed mass and energy and different radii. The lightcurves
begin to differ in the very early phase for both models, especially
for RW11, which begin to differ within one day after explosion. The
magnitude for a smaller radius begins to drop sooner than that for a
larger radius in both models.

\begin{figure}[!ht]
   \centering
   \includegraphics[angle=90,width=7.0cm]{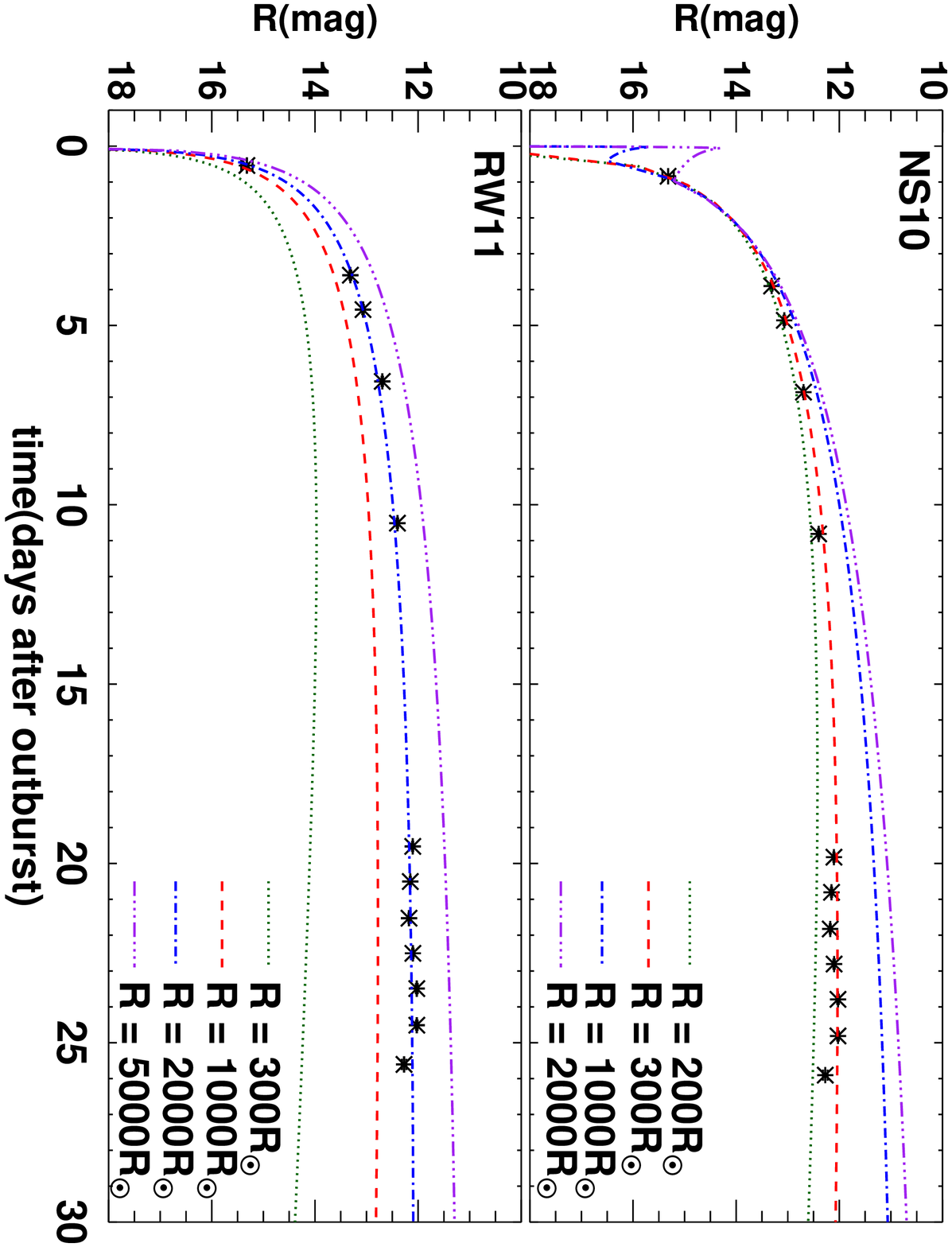}
      \caption{Zoom in of the optical lightcurves of SN 1998S in comparison to the analytical model
      of NS10(top panel with M = 15$M_{\bigodot}$ and E = 8.75$\times$10$^{51}$
      ergs) and RW11(bottom panel with M = 15$M_{\bigodot}$ and E =
      3.0$\times$10$^{51}$) with different radii. The dates of explosion for NS10's
      and RW11's models in the figure are JD 2450874.34 and JD 2450874.64 respectively.
              }
         \label{FigVibStab}
   \end{figure}

   In figure 6, we show the lightcurves for NS10 and RW11 with fixed radius and energy and varying
   mass. A smaller mass results in
brighter magnitudes for both models. The different lightcurves show
similar tendency. They mainly differ in magnitudes. By adding a
constant, all lightcurves can fit the data.

\begin{figure}[!ht]
   \centering
   \includegraphics[angle=90,width=7.0cm]{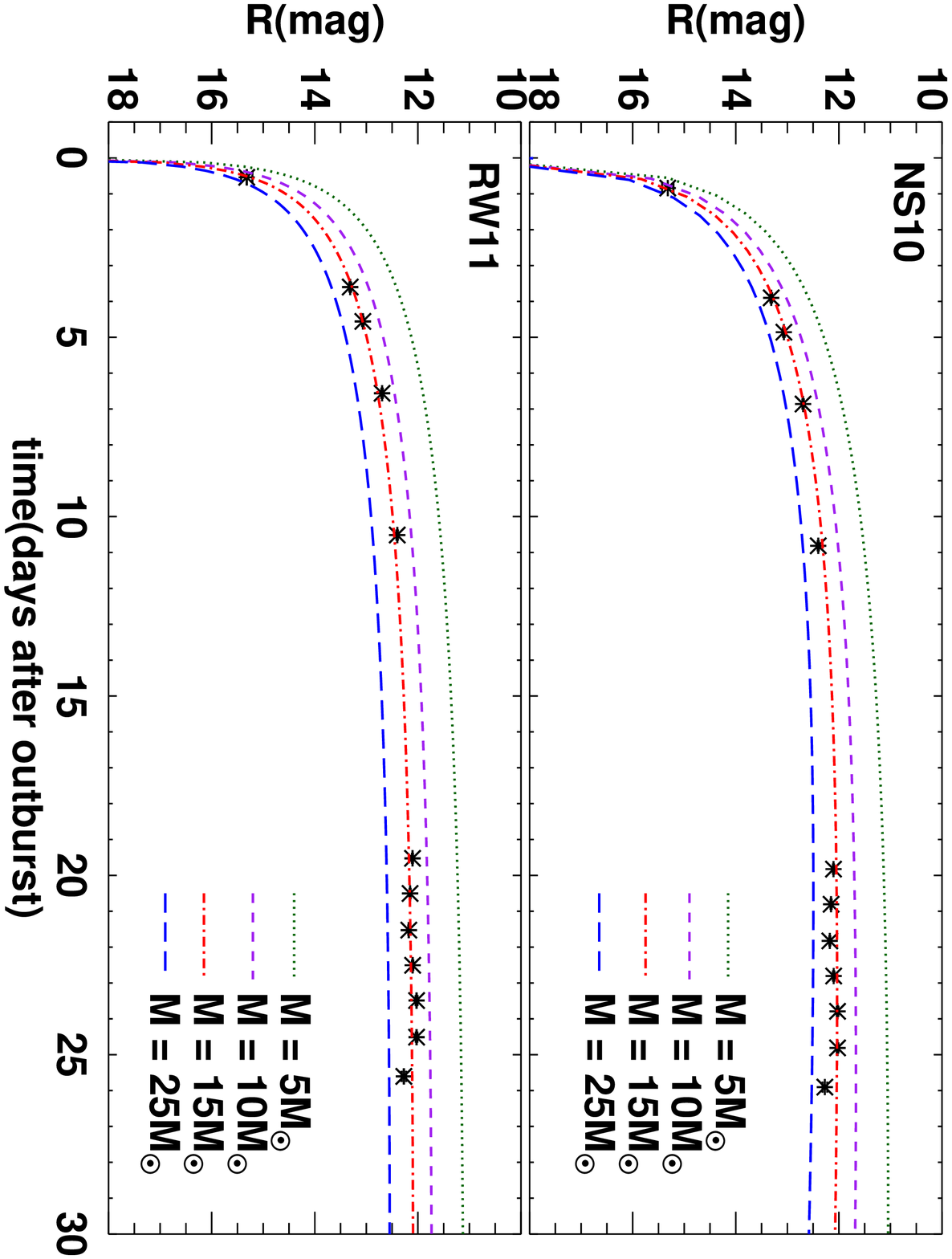}
      \caption{Zoom in of the optical lightcurves of SN 1998S in comparison to the analytical model
      of NS10(top panel with R = 300$R_{\bigodot}$ and E = 8.75$\times$10$^{51}$
      ergs) and RW11(bottom panel with R = 2000$R_{\bigodot}$ and E =
      3.0$\times$10$^{51}$ ergs) with different masses. The dates of explosion for NS10
      and RW11 in the figure are JD 2450874.34 and JD 2450874.64 respectively.
              }
         \label{FigVibStab}
   \end{figure}

We plot the lightcurves for both models with fixed mass and radius
and different energies in figure 7. In RW11, a small change in
energy results in larger differences in magnitudes than NS10, but
different lightcurves show similar trends and can fit the data
 by adding a constant.

\begin{figure}[!ht]
   \centering
   \includegraphics[angle=90,width=7cm]{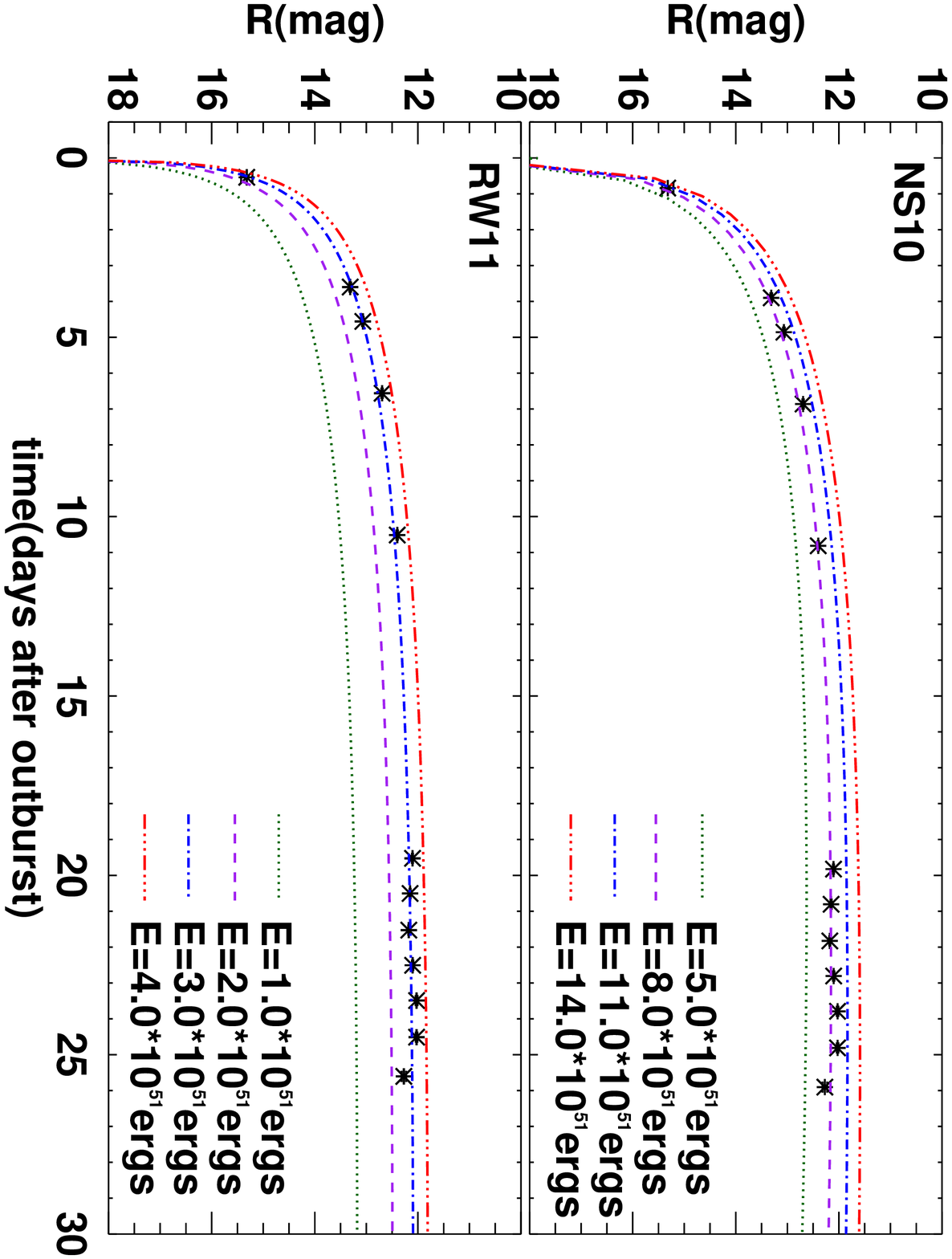}
      \caption{Zoom in of the optical lightcurves of SN 1998S in comparison to the analytical model
      of NS10(top panel with R = 300$R_{\bigodot}$ and M = 15$M_{\bigodot}$
      ergs) and RW11(bottom panel with R = 2000$R_{\bigodot}$ and M =
      15$M_{\bigodot}$) with different energies. The dates of explosion for NS10
      and RW11 in the figure are JD 2450874.34 and JD 2450874.64 respectively.
              }
         \label{FigVibStab}
   \end{figure}

From the previous results, radius seems to have a more important
impact in the shape of lightcurves. In figure 8, we try to fit our
data using both models, each with the best fitting radius, different
masses and the best corresponding energy. In both models, a bigger
mass requires a bigger energy. With suitable combination of mass and
energy, similar lightcurves can be produced. For the mass of 5 to 25
$M_{\bigodot}$ , the corresponding energy range which can fit the
data is 3.5$\times$10$^{51}$ to 13.5$\times$10$^{51}$ ergs for NS10
and 1.25$\times$10$^{51}$ to 4.5$\times$10$^{51}$ ergs for RW11.
These ranges are reasonable for a red supergiant. Both NS10 and RW11
can constrain the radius but cannot distinguish between different
combinations of mass and energy. We note that for RW11, the
lightcurves for R = 2000$R_{\bigodot}$, M = 5$M_{\bigodot}$ and E =
1.25$\times$10$^{51}$ ergs can fit our data well enough. This
modelling result is consistent with Chugai(2001) in which he assumed
R = 2.4$\times$10$^{14}$ cm($\sim$ 3500$R_{\bigodot}$), M =
5$M_{\bigodot}$ and E = 1.1$\times$10$^{51}$ ergs.

 \begin{figure}[!ht]
   \centering
   \includegraphics[angle=90,width=7cm]{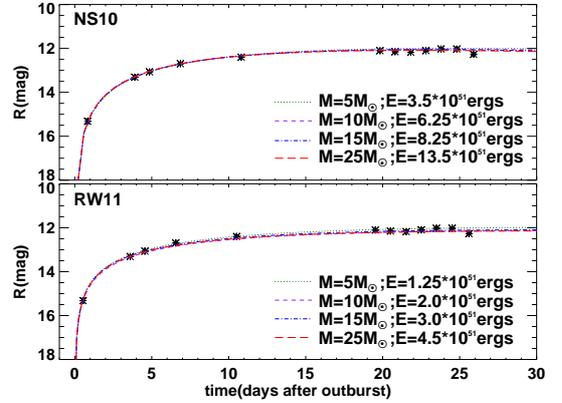}
      \caption{Zoom in of the optical lightcurves of SN 1998S in comparison to
the analytical model of NS10(top panel with R = 300$R_{\bigodot}$)
and RW11(bottom panel with R = 2000$R_{\bigodot}$). For each model,
the lightcurves for different masses and the best corresponding
energy are shown. The dates of explosion for NS10
      and RW11 in the figure are JD 2450874.34 and JD 2450874.64 respectively.
              }
         \label{FigVibStab}
   \end{figure}

We try to compare the temperatures derived from both models to that
of SN 1998S, which are estimated by blackbody fit. In figure 1, we
plot the temperatures for both models using the best fitting
parameters derived in figure 4. Both models predict a lower
temperature than the data. We also try to fit the temperatures using
different sets of parameters. The best fit for NS10 predicts a
temperature of $\sim$ 67000K on the day of discovery, which amounts
to $\sim$ 0.011 in the term i$_{3}$(V$_{SN}$ - I$_{SN}$) of
equation(2). For RW11, the temperature found for this day is $\sim$
55000K. This also amounts to $\sim$ 0.011 for the above-mentioned
term. Since this value differs from the error we quote by a
magnitude of 0.001, even if the temperature is really that high, the
results are still not affected.


\section{Discussion and Conclusion}

We present early photometry of SN 1998S in this work and use
analytical models to fit our data. The dates of explosion derived
from both models agree with each other very well, only with a small
difference of 0.3 day. For NS10, the date of explosion is 1998 March
1.84(JD 2450874.34) and it is 1998 March 2.14(JD 2450874.64) for
RW11. In NS10, the lightcurves show a bump in the very early phase
which is the transition between the planar and the spherical phase.
This bump indicates the magnitude during transition, which is an
important constraint to the parameters. At the very early phase
after supernova explosion, the energy concentrates on the high
energy band. If our first data point was taken in the \emph{U} band
instead of unfiltered, we may be able to better constrain the
parameters of the supernova. GALEX detected UV emission($<$1 day)
from SNe SNLS-04D2dc and SNLS-06D1jd(Gezari et al. 2008). The
authors model the very early lightcurves up to 55 hours and find the
mass loss to be $\sim$ 10$^{-3}$$M_{\bigodot}$ yr$^{-1}$. We took
four data points during the first 10 days, with the first one taken
within one day after explosion. This sampling frequency is dense
enough and is significant in determining the parameters.

Both NS10 and RW11 for an RSG can fit the early lightcurves well. In
NS10, the breakout is observed and then followed by a broken
power-law decay of the luminosity in the planar phase. Then in the
spherical phase, the luminosity rises again until the frequency
range peaks in the blackbody spectrum. The transition between the
planar and spherical phases causes the early bump in the analytical
lightcurves. This bump is not observed in RW11 since this model has
got only one phase. In the upper panel of fig.4(NS10), a larger
radius results in a longer planar phase and a brighter magnitude at
the transition. The duration of the planar phase and the magnitude
at transition are important in constraining the model parameters.
However, this very early phase is not considered in RW11 and the
result may be different if the breakout phase is considered in this
model. From the results of the fit, we find the radius of the
progenitor to be $\sim$ 300$R_{\bigodot}$(NS10) and $\sim$
2000$R_{\bigodot}$(RW11). The later result is consistent with
Chugai(2001). The ranges of mass and energy found from both models
are within that of a red supergiant. These two models derive quite
different parameters.

The temperatures derived from both models cannot fit the data well.
We can only obtain a satisfactory fit with different sets of
parameters for both models. The derived temperatures would be a lot
higher than our extrapolated result. However, this would only
contribute to a small change of a magnitude of $\sim$ 0.001 in the
error. The final result is not really affected.

\begin{acknowledgements}
 The authors thank E. Nakar and J. S. Deng for helpful
 discussions.
\end{acknowledgements}

\clearpage

\begin{table}[!ht]
\caption{Standard Magnitudes and Temperatures of SN 1998S(t$_{exp}$
= JD 2450874.49)} \label{table:1}
\centering                                    
\begin{tabular}{c c c c c c}        
\hline\hline                 
 Day & Julian Day(2450000+) &\emph{B} band & \emph{V} Band & \emph{R} Band & Temperature(K) \\    
\hline                        

 0.69  &       875.18 &          -       &      -          &  15.32 (0.11)  &      -        \\
3.75   &       878.24 &   13.06 (0.06)   &   13.29 (0.01)  &  13.31 (0.02)  &  23000 (7500) \\
4.71   &       879.20 &   12.72 (0.06)   &   13.08 (0.01)  &  13.07 (0.02)  &  22000 (6500) \\
6.71   &       881.20 &   12.60 (0.05)   &   12.74 (0.01)  &  12.70 (0.01)  &  21000 (6000) \\
10.66  &       885.15 &   12.32 (0.05)   &   12.39 (0.01)  &  12.40 (0.01)  &  16000 (4100) \\
19.68  &       894.17 &   12.38 (0.04)   &   12.31 (0.01)  &  12.11 (0.01)  &  13000 (2800) \\
20.66  &       895.15 &   12.51 (0.03)   &   12.28 (0.01)  &  12.15 (0.01)  &  12000 (2400) \\
21.68  &       896.17 &   12.45 (0.04)   &   12.34 (0.01)  &  12.18 (0.01)  &  11000 (2200) \\
22.66  &       897.15 &   12.64 (0.04)   &   12.45 (0.02)  &  12.10 (0.01)  &  11000 (6000) \\
23.64  &       898.13 &   13.00 (0.04)   &   12.48 (0.01)  &  12.03 (0.02)  &  10000 (2000)  \\
24.66  &       899.15 &   12.92 (0.04)   &   12.44 (0.02)  &  12.03 (0.02)  &  10000 (1800)  \\
25.75  &       900.24 &   12.84 (0.14)   &   12.68 (0.05)  &  12.27 (0.45)  &  9700 (2600)   \\

%
\hline
\end{tabular}
\end{table}

\begin{table}[!ht]
\caption{Residuals of the models of NS10 and RW11 using different
parameters} \label{table:2}
\centering                                    
\begin{tabular}{c|cccc|cccc}        
\hline\hline                 
     & &   NS10 & &  &    &  RW10 & &  \\

 Julian Day(2450000+) & modal 1 & modal 2 & modal 3 & modal 4 & modal 1 & modal 2 & modal 3 & modal 4 \\    
\hline                        
    875.18 &   -0.05   &    0.02    &    0.15    &    0.22     & -0.15 &   -0.07  &  0.00    &   0.00    \\
  878.24 &   -0.02   &   -0.01    &    0.03    &    0.01     & -0.04 &   -0.04  & -0.03    &  -0.06    \\
  879.20 &   -0.01   &   -0.01    &    0.01    &   -0.04     &  0.01 &    0.00  & -0.00    &  -0.05    \\
  881.20 &    0.05   &    0.01    &   -0.01    &   -0.10     &  0.12 &    0.08  &  0.04    &  -0.00    \\
  885.15 &    0.02   &   -0.09    &   -0.15    &   -0.32     &  0.14 &    0.06  & -0.02    &  -0.09    \\
  894.17 &    0.15   &   -0.03    &   -0.26    &   -0.58     &  0.24 &    0.07  & -0.09    &  -0.19    \\
  895.15 &    0.10   &   -0.08    &   -0.33    &   -0.66     &  0.18 &    0.01  & -0.16    &  -0.26    \\
  896.17 &    0.08   &   -0.11    &   -0.37    &   -0.72     &  0.16 &   -0.03  & -0.20    &  -0.31    \\
  897.15 &    0.16   &   -0.04    &   -0.32    &   -0.68     &  0.23 &    0.03  & -0.14    &  -0.26    \\
  898.13 &    0.25   &    0.04    &   -0.25    &   -0.63     &  0.31 &    0.10  & -0.08    &  -0.20     \\
  899.15 &    0.26   &    0.04    &   -0.26    &   -0.66     &  0.31 &    0.09  & -0.10    &  -0.22     \\
  900.24 &    0.03   &   -0.20    &   -0.52    &   -0.93     &  0.06 &   -0.16  & -0.36    &  -0.49     \\

\hline
\end{tabular}
NS10:

Modal 1: R = 200$R_{\bigodot}$; M = 5$M_{\bigodot}$; E =
4.0$\times$10$^{51}$ ergs

Modal 2: R = 300$R_{\bigodot}$; M = 15$M_{\bigodot}$; E =
8.75$\times$10$^{51}$ ergs

Modal 3: R = 500$R_{\bigodot}$; M = 15$M_{\bigodot}$; E =
7.5$\times$10$^{51}$ ergs

Modal 4: R = 1000$R_{\bigodot}$; M = 20$M_{\bigodot}$; E =
9.0$\times$10$^{51}$ ergs

RW10:

Modal 1: R = 1000$R_{\bigodot}$; M = 5$M_{\bigodot}$; E =
1.5$\times$10$^{51}$ ergs

Modal 2: R = 2000$R_{\bigodot}$; M = 15$M_{\bigodot}$; E =
3.0$\times$10$^{51}$ ergs

Modal 3: R = 3500$R_{\bigodot}$; M = 15$M_{\bigodot}$; E =
2.5$\times$10$^{51}$ ergs

Modal 4: R = 5000$R_{\bigodot}$; M = 25$M_{\bigodot}$; E =
3.0$\times$10$^{51}$ ergs

\end{table}

\end{document}